\documentclass[prl,twocolumn,showpacs,preprintnumbers,amsmath,amssymb]{revtex4}

\usepackage{mathrsfs}
\usepackage{graphicx}
\usepackage{dcolumn}
\usepackage{bm}

\begin{document}

\title{Information Thermodynamics on Causal Networks}

\author{Sosuke Ito$^1$ and Takahiro Sagawa$^{2}$}

\affiliation{$^1$Department of Physics, The University of Tokyo, 7-3-1 Hongo, Bunkyo-ku, Tokyo 113-0033, Japan\\
$^2$Department of Basic Science, The University of Tokyo, 3-8-1 Komaba, Meguro-ku, Tokyo 153-8902, Japan}
\date{\today}

\begin{abstract}
We study nonequilibrium thermodynamics of complex information flows induced by interactions between multiple fluctuating systems. Characterizing nonequilibrium dynamics by causal networks (i.e., Bayesian networks), we obtain novel generalizations of the second law of thermodynamics and the fluctuation theorem, which include an informational quantity characterized by the topology of the causal network. Our result implies that the entropy production in a single system in the presence of multiple other systems is bounded by the information flow between these systems. We demonstrate our general result by a simple model of biochemical adaptation.
\end{abstract}


\maketitle

\textit{Introduction.}---Nonequilibrium equalities for small thermodynamic systems such as molecular motors have been intensively investigated in the last two decades~\cite{Seifert, Sekimoto}. The second law of thermodynamics can be derived from the Jarzynski equality~\cite{Jarzynski} and the fluctuation theorems (FTs)~\cite{Evans,Crooks,Lebowitz,Jarzynski2,Evans2}. The second law is expressed in terms of the ensemble average of the entropy production $\sigma$:
\begin{equation}
\left< \sigma \right> \geq 0,
\label{second law}
\end{equation}
where $\left<{\dots}\right>$ describes the ensemble average.
We note that $\sigma$ reduces to the difference in the free-energy change $\Delta{F}$ and the work $W$ performed on the system such that $\sigma={\beta}({W}-{\Delta}F)$, when the system is attached to a single heat bath with inverse temperature $\beta$, and the initial and final states are in thermal equilibrium.

On the other hand, in the presence of feedback control by Maxwell's demon~\cite{Maxwell, Szilard, Maxwell2}, the second law seems to be violated; i.e., $\left<{\sigma}\right>$ can be negative. For such cases, the second law has been generalized as
\begin{equation}
\left< \sigma \right> \geq \left< \Delta I\right>,
\label{information thermodynamics}
\end{equation}
where $\left<\Delta{I}\right>$ is the mutual information that is exchanged between the system and the demon~\cite{Sagawa,Sagawa2}. Such a Maxwell's demon has been experimentally demonstrated with a colloidal particle~\cite{Toyabe}. While the relationship between information and thermodynamics has been studied in several simple setups with the demon~\cite{Landauer,Sagawaq,Touchette,CaoDinis,Touchette2,KimQian,CaoFeito,CaoFeito2,Fujitani,Horowitz,Ponmurugan,Morikuni,Kim,Ito,Horowitz2,Abreu,Vaikuntanathan,Horowitz3,Granger,Esposito,Sagawa3,Munakata,Esposito2,Abreu3,Lahiri,Sagawa4,Cao,Horowitz4,Kundu,Bauer,Mandal,Still,Barato2,Munakata2,Barato, Esposito3,Mandal2,Cardoso,Berut}, the general theory has been elusive for more complex cases in which multiple systems exchange information many times.

	In this Letter, we derive a novel nonequilibrium equality in the presence of complex information flows between multiple stochastic systems. Our result involves a new informational term that is characterized by the topology of the causal structure of the dynamics. The informational quantity consists of the initial correlation between the target system and other systems, the information transfer from the system to others during the dynamics, and the final correlation between them. Our result can reproduce inequality (\ref{information thermodynamics}) for special cases. In order to describe nonequilibrium dynamics of multiple systems, we use Bayesian networks (BNs)~\cite{Bayesian} that topologically represent the causal structure of the dynamics.
	
	Our theory is applicable to quite a broad class of nonequilibrium dynamics such as an information transfer between multiple Brownian particles and information processing in autonomous nanomachines. We illustrate our result by a chemical model of biological adaptation with time-delayed feedback. Our result implies that information processing plays a crucial role in biochemical reactions.
	
\textit{Bayesian networks.}---First,
			\begin{figure}
		\centering
		\includegraphics[width=85mm]{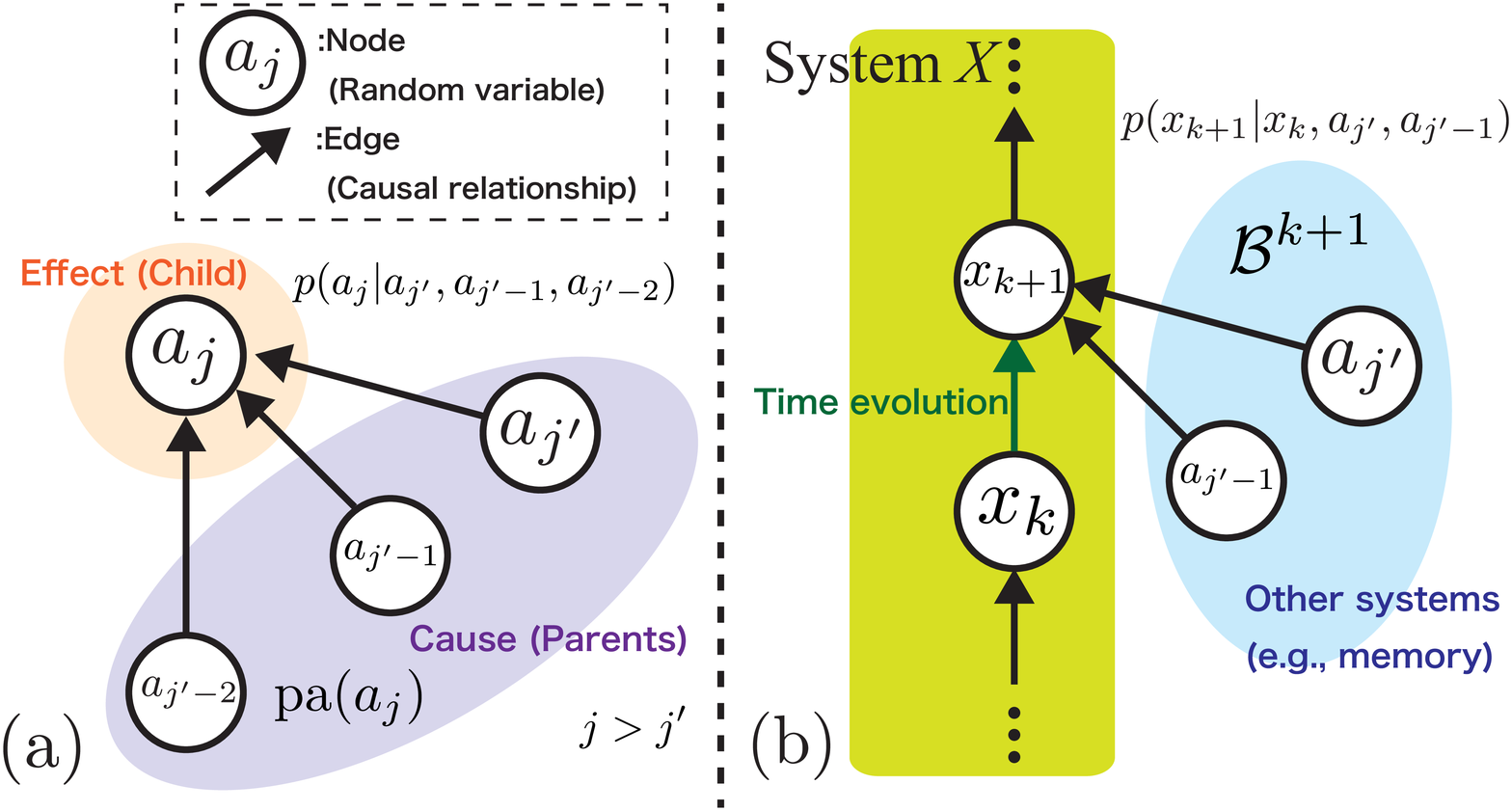}
		\caption{(color online). (a) Schematic of a BN. (b) Stochastic dynamics of system $X$ under the influence of other systems.}
		\label{fig:bayesianshematic}
	\end{figure}
 we briefly discuss the basic concepts of BNs [see also Fig.~\ref{fig:bayesianshematic}(a)]. Let $\mathcal{A}=\{a_j|j=1,2,\dots,N_{\mathcal{A}} \}$ be the set of random variables that are associated with the nodes of a BN, where $N_{\mathcal{A}}$ is the number of the nodes. When an edge $a_{j'}{\rightarrow}a_j$ exists, there is a causal relationship from $a_{j'}$ to $a_{j}$, where we say that $a_{j'}$ is a parent of $a_j$. We denote by ${\rm pa}(a_j)$ the set of parents of $a_j$. Here, the order of $a_1,a_2,\dots$ is determined by the causal relationship in the BN such that $a_{j}$ cannot be a parent of $a_{j'}$ if $j'<j$. This order is referred to as the topological ordering. We characterize stochastic dynamics in the BN by the conditional probability $p(a_j|a_{j-1},\dots,a_1) = p(a_j|{\rm pa}(a_j))$ that describes the probability of  $a_j$ under the condition of a particular realization of ${\rm pa}(a_j)$.  We write $p(a_j|\emptyset)=p(a_j)$, where $\emptyset$ is the empty set. Because of the chain rule in the probability theory, we obtain the joint probability distribution of the all random variables~\cite{Bayesian}:
\begin{equation}
p(\mathcal{A}) = \prod_{j=1}^{N_{\mathcal A}} p(a_j|{\rm pa}(a_j)).
\label{definition of joint probability}
\end{equation}
The ensemble average of the arbitrary function $g(\mathcal{A})$ is defined as $\left<g\right>\equiv{\sum}_{\mathcal{A}}p(\mathcal{A})g(\mathcal{A})$.

				\begin{figure}
		\centering
		\includegraphics[width=85mm]{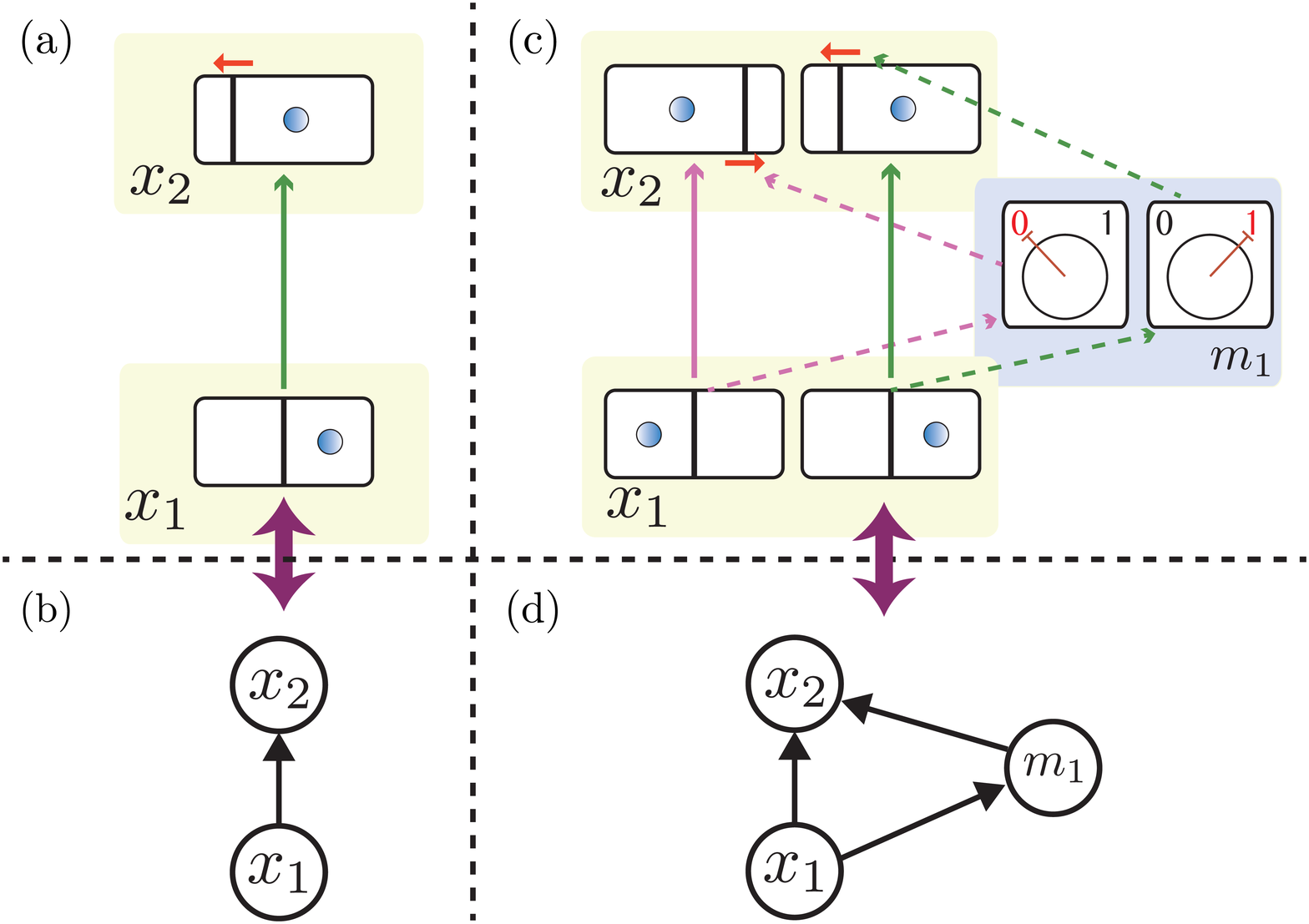}
		\caption{(color online). (a) Time evolution of a single-molecule gas without feedback control. (b) BN corresponding to (a). (c) The Szilard engine with feedback control by a memory device. (d) BN corresponding to (c). }
		\label{fig:bayesian}
	\end{figure}
	We next describe how we use BNs to describe stochastic dynamics [see also Fig.~\ref{fig:bayesianshematic}(b)]. We consider a situation in which system $X$ interacts with other systems.  The probability distribution of all the systems for the entire process is given by Eq. (\ref{definition of joint probability}), where $a_j$ corresponds to a state of a system at a particular time. $\mathcal{A}$ consists of all states in the time evolution of both system $X$ and other systems.
	
	We also use the notation $X$ to describe the time evolution of system $X$; we write $X\equiv\{x_k|k =1,2,\dots,N\}$ ($\subseteq{\mathcal{A}}$), where $x_k$ is the state of system $X$ at time $k$, and $\subseteq$ is the symbol of the subset. We assume that $x_{k}$ is a parent of $x_{k+1}$. We also assume that $x_k$ cannot be a parent of $x_{k'}$ for $k'\neq{k+1}$. We note that the time evolution of $X$ is characterized by the chain $x_1\to{x_2}\to{\cdots}\to{x_N}$. 
	
	For instance, Fig.~\ref{fig:bayesian}(a) shows an expansion of a single-molecule gas, which can be described by the BN shown in Fig.~\ref{fig:bayesian}(b). This BN shows the time evolution such that $p(x_1{,}x_2){=}p(x_2{|}x_1){p(x_1)}$, where $x_1$ and $x_2$, respectively, describe the initial and final positions of the particle. In Fig.~\ref{fig:bayesian}(c), we illustrate the Szilard engine~\cite{Szilard} that is a standard model of Maxwell's demon. Figure~\ref{fig:bayesian}(d) shows the corresponding BN, where $m_1$ describes a memory state that is correlated with $x_1$. This BN shows the time evolution of the total system $p(x_1,{x_2},{m_1})=p(x_2|x_1,{m_1})p(m_1|x_1)p(x_1)$.

\textit{Entropy production and mutual information.}---We introduce the entropy production in stochastic thermodynamics in terms of the BN. We assume that system $X$ is coupled to heat baths with inverse temperatures $\beta_\alpha$ $(\alpha=1,2,\dots,n_{\rm bath})$. Let $Q_\alpha$ be the heat absorbed by $X$ from the $\alpha$th bath. Because of the standard definition in stochastic thermodynamics~\cite{Seifert}, the entropy production in $X$ is given by $\sigma{\equiv}\Delta{s}_{\rm bath}+\ln{p}(x_1)-\ln{p}(x_N)$, where $x_{1}$ ($x_{N}$) is the initial (final) state of $X$ and $\Delta{s}_{\rm bath}{\equiv}-\sum_\alpha\beta_\alpha{Q}_\alpha$ is the entropy change in the baths. Let $\Delta{s}_{\rm bath}^{k+1}$ be the entropy change in the baths from time $k$ to $k+1$ such that $\Delta{s}_{\rm bath}=\sum_{k=1}^{N-1}\Delta{s}_{\rm bath}^{k+1}$. In quite a broad class of nonequilibrium dynamics including multidimensional Langevin dynamics (see the Supplemental Material), $\Delta{s}_{\rm bath}^{k+1}$ satisfies the detailed FT~\cite{Jarzynski2, Evans2}:
\begin{equation}
\Delta s_{\rm bath}^{k+1} \equiv \ln \frac{p\left( x_{k+1} \left|x_k ,\mathcal{B}^{k+1} \right) \right. }{p_B\left( x_{k} \left| x_{k+1}, \mathcal{B}^{k+1} \right) \right.},
\label{definition of heat}
\end{equation}
where $\mathcal{B}^{k+1}$ is defined as $\mathcal{B}^{k+1}\equiv{\rm pa}(x_{k+1})\setminus\{x_k\}$ with $\setminus$ indicating the relative complement of two sets. $\mathcal{B}^{k+1}$ means the set of random variables which affect the time evolution of $X$ from states $x_k$ to $x_{k+1}$ [see also Fig.~\ref{fig:bayesianshematic}(b)]. $p_B$ describes the probability distribution of backward paths. 

We next introduce mutual information that plays a crucial role in this study. Let $\mathcal{A}_1$, $\mathcal{A}_2$ and $\mathcal{A}_3$ be arbitrary sets of random variables. We define
$I(\mathcal{A}_1:\mathcal{A}_2|\mathcal{A}_3)\equiv\ln{p(\mathcal{A}_1,\mathcal{A}_2|\mathcal{A}_3)}-\ln{p(\mathcal{A}_1|\mathcal{A}_3)}-\ln{p(\mathcal{A}_2|\mathcal{A}_3)}$,
where we write $I(\mathcal{A}_1:\mathcal{A}_2|\mathcal{A}_3=\emptyset)=I(\mathcal{A}_1:\mathcal{A}_2)$. Its ensemble average $\left<I(\mathcal{A}_1:\mathcal{A}_2|\mathcal{A}_3)\right>$ is the mutual information between $\mathcal{A}_1$ and $\mathcal{A}_2$ under the condition of $\mathcal{A}_3$.
	
\begin{figure}
		\centering
		\includegraphics[width=80mm]{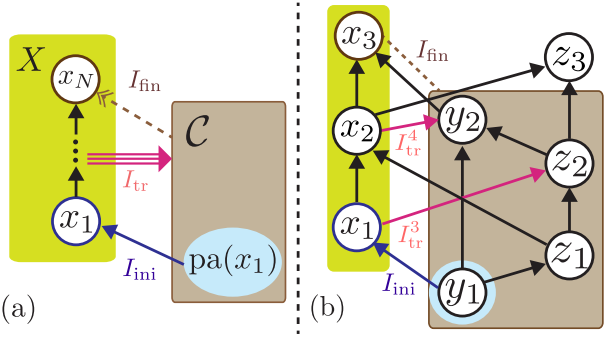}
		\caption{(color online). (a) Schematic of $\mathcal{C}$ and ${\rm pa}(x_1)$. $\mathcal{C}$ is the history of other systems that can affect the final state $x_N$. ${\rm pa}(x_1)$ describes the variables correlated with the initial state $x_1$. (b) An example of BN that describes three-body interactions.}
		\label{fig:bayesianexample}
	\end{figure}

\textit{Main result.}---In order to discuss the main result, we introduce set $\mathcal{C}\equiv\{a_1,a_2,\dots,a_J\}\setminus{X}$, where $a_J$ is chosen to satisfy $a_{J}=x_N$ [see also Fig.~\ref{fig:bayesianexample}(a)]. Here, $\mathcal{C}$ is the history of the other systems that can affect the final state $x_N$. We denote the elements of $\mathcal{C}$ as $\mathcal{C}=\{c_l|l=1,2,\dots,N'\}$, where $c_1,c_2,\dots$ are in the topological ordering.
		
	We now state the main result of this Letter. In the foregoing setup, we have a new generalization of the integral FT (IFT):
\begin{equation}
\left< \exp\left[-\sigma +  \Theta \right] \right> = 1.
\label{bayesian jarzynski equality}
\end{equation}
Here, the key quantity $\Theta$ is the informational quantity characterized by the topology of the BN:
\begin{align}
\Theta &\equiv I_{\rm fin}  - I_{\rm ini}- \sum_{l=1}^{N'}  I_{\rm tr}^{l},
\label{partition}\\
I_{\rm fin} &\equiv I (x_N : \mathcal{C}),
\label{fin mutual information} \\
I_{\rm ini} &\equiv I(x_1: {\rm pa}(x_1)),
\label{ini mutual information} \\
I_{\rm tr}^{l} &\equiv I (c_{l} :{\rm pa}_X(c_{l}) | \mathcal{C}_{l-1}),
\label{transfer entropy}
\end{align}
where $\mathcal{C}_{l-1}\equiv\{c_{l'}|l'=1,2,\dots,l-1\}$ and ${\rm pa}_X(a_j)\equiv{\rm pa}(a_j)\cap{X}$, with $\cap$ indicating the intersection. Here, $I_{\rm ini}$ characterizes the initial correlation between $X$ and the other systems, while $I_{\rm fin}$ characterizes the final correlation that remains at the end of the dynamics. On the other hand, $I_{\rm tr}^l$ is the transfer entropy~\cite{Schreiber} that characterizes the information transfer into $c_l$ from $X$ during the dynamics (see the Supplemental Material). For example, in the case of Fig.~\ref{fig:bayesianexample}~(b), we obtain $I_{\rm fin}=I(x_3:\{y_1,z_1,z_2,y_2\})$, $I_{\rm ini}=I(x_1:y_1)$, $I_{\rm tr}^{1}=I_{\rm tr}^{2}=0$, $I_{\rm tr}^{3}=I(z_2:x_1|y_1,z_1)$ and $I_{\rm tr}^{4}=I(y_2:x_2|y_1,z_1,z_2)$. We will discuss the proof of Eq.~(\ref{bayesian jarzynski equality}) later.

By using the Jensen inequality for convex functions, i.e., $\langle\exp[g]\rangle\geq\exp[\langle{g}\rangle]$, we obtain
\begin{equation}
\left< \sigma \right> \geq \left<I_{\rm fin} \right>- \left<I_{\rm ini} \right> - \sum_{l=1}^{N'} \left< I^{l}_{\rm tr} \right>,
\label{bayesian information thermodynamics}
\end{equation}
which is a novel generalization of the second law of thermodynamics for subsystem $X$ in the presence of complex information flows.

In the following, we illustrate that our main result (\ref{bayesian jarzynski equality}) can reproduce known nonequilibrium relations for special cases in a unified way, and moreover, can lead to new generalizations of the IFT.

\textit{Example 1.}---We consider
	\begin{figure}
		\centering
		\includegraphics[width=80mm]{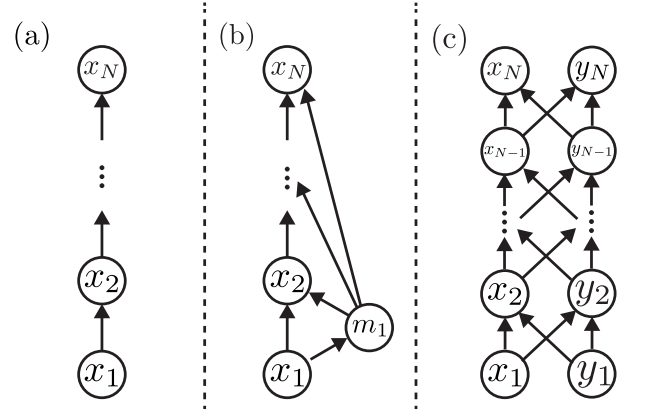}
		\caption{(a) BN corresponding to a simple Markov chain. (b) BN corresponding to feedback control. (c) BN corresponding to two Brownian particles.}
		\label{fig:bayesian2}
	\end{figure}
	 the Markov chain shown in Fig.~\ref{fig:bayesian2}(a). We have $\mathcal{C}=\emptyset$ and ${\rm pa}(x_1)=\emptyset$, and therefore $I_{\rm fin}=0$, $I_{\rm ini}=0$, and $\Theta=0$. We then reproduce the conventional IFT: $\left<\exp[-\sigma]\right>=1$, which leads to inequality (\ref{second law}).

\textit{Example 2.}---We next consider a system with feedback control shown in Fig.~\ref{fig:bayesian2}(b),  where $m_1$ describes a state of the memory.  State $x_1$ is measured by the memory, and the outcome $m_1$ is used for the feedback control. We have $\mathcal{C}=\{m_1\}$ and ${\rm pa}(x_1)=\emptyset$, and therefore $I_{\rm fin}= I(x_{N}:m_1)$, $I_{\rm ini}=0$, $I^{1}_{\rm tr}=I(x_{1}: m_1)$, and $\Theta=I(x_{N} :m_1)-I(x_{1} :m_1)$. We then reproduce a generalized IFT obtained in Ref.~\cite{Sagawa2}: $\left<\exp[-\sigma+\Delta{I}]\right>=1$, which leads to inequality (\ref{information thermodynamics}). We note that in the case of the discrete repeated feedback, a previous result~\cite{Horowitz} can be derived from Eqs. (\ref{bayesian jarzynski equality}) and (\ref{bayesian information thermodynamics}) (see the Supplemental Material).

\textit{Example 3.}---We next consider the two-dimensional Langevin equation that describes an interaction between two Brownian particles:
\begin{align}
\gamma^x \frac{d x}{dt}(t) &= f^x (x(t), y(t)) +\xi^x(t), 
\label{Langevin x}\\
\gamma^y \frac{d y}{dt}(t) &= f^y (x(t), y(t)) +\xi^y(t),
\label{Langevin y}
\end{align}
where $t$ is time, $\gamma^x$ and $\gamma^y$ are friction coefficients, $f^x$ and $f^y$ are mechanical forces, and $\xi^x$ and $\xi^y$ are independent white-Gaussian noises with variances $2\gamma^x/\beta^x$ and $2\gamma^y/\beta^y$, respectively. Let $\Delta{t}$ be an infinitesimal time interval. We discretize the dynamics as $x_k\equiv{x}(t=k\Delta{t})$ and $y_k\equiv{y}(t=k\Delta{t})$, and introduce the corresponding BN by Fig.~\ref{fig:bayesian2} (c) where system $X$ corresponds to one particle with coordinate $x(t)$. We then have $\mathcal{C}=\{y_1,\dots,y_{N-1}\}$ and ${\rm pa}(x_1)=\emptyset$, and therefore $I_{\rm fin}=I(x_{N}:\{y_1,\dots,y_{N-1}\})$, $I_{\rm ini}=0$, $I_{\rm tr}^{l}=I(y_l:x_{l-1}|y_{l-1},\dots,y_1)$ [$I^1_{\rm tr}=I(y_1:\emptyset)=0$], and $\Theta=I_{\rm fin}-\sum_{l=1}^{N-1}I^{l}_{\rm tr}$. We note that $\Delta{s}_{\rm bath}=-\beta^xQ_x$, where
$Q_x$ is the heat absorbed by system $X$ from the bath~\cite{Sekimoto2} (see the Supplemental Material for details).

In this case, inequality (\ref{bayesian information thermodynamics}) implies that the entropy production of one particle is bounded by the information flow into the other particle and the final correlation with it.  As shown in the Supplemental Material, such a result is valid for multidimensional cases, in general, which enables us to characterize the entropy production in one particle that interacts with multiple particles in terms of information exchanges between them. We note that the entropy production in a single particle of a multidimensional Langevin system is closely related to experiments on the role of the hidden degrees of freedom~\cite{Mehl, Speck}.
	
\textit{Model of biological adaptation.}---We next discuss an application of our general result to a biochemical system. The significance of information processing in biochemical networks has been presented, for example, in Refs.~\cite{Tostevin,Mehta,Cheong}. In particular, feedback control plays a key role in biological adaptations such as bacterial chemotaxis~\cite{Tu, Lan}. We show that the free-energy difference is bounded by an informational quantity in the presence of a chemical feedback loop in a simple model of adaptation with the time-delay effect~\cite{Brandman}. 
	
				\begin{figure}
		\centering
		\includegraphics[width=85mm]{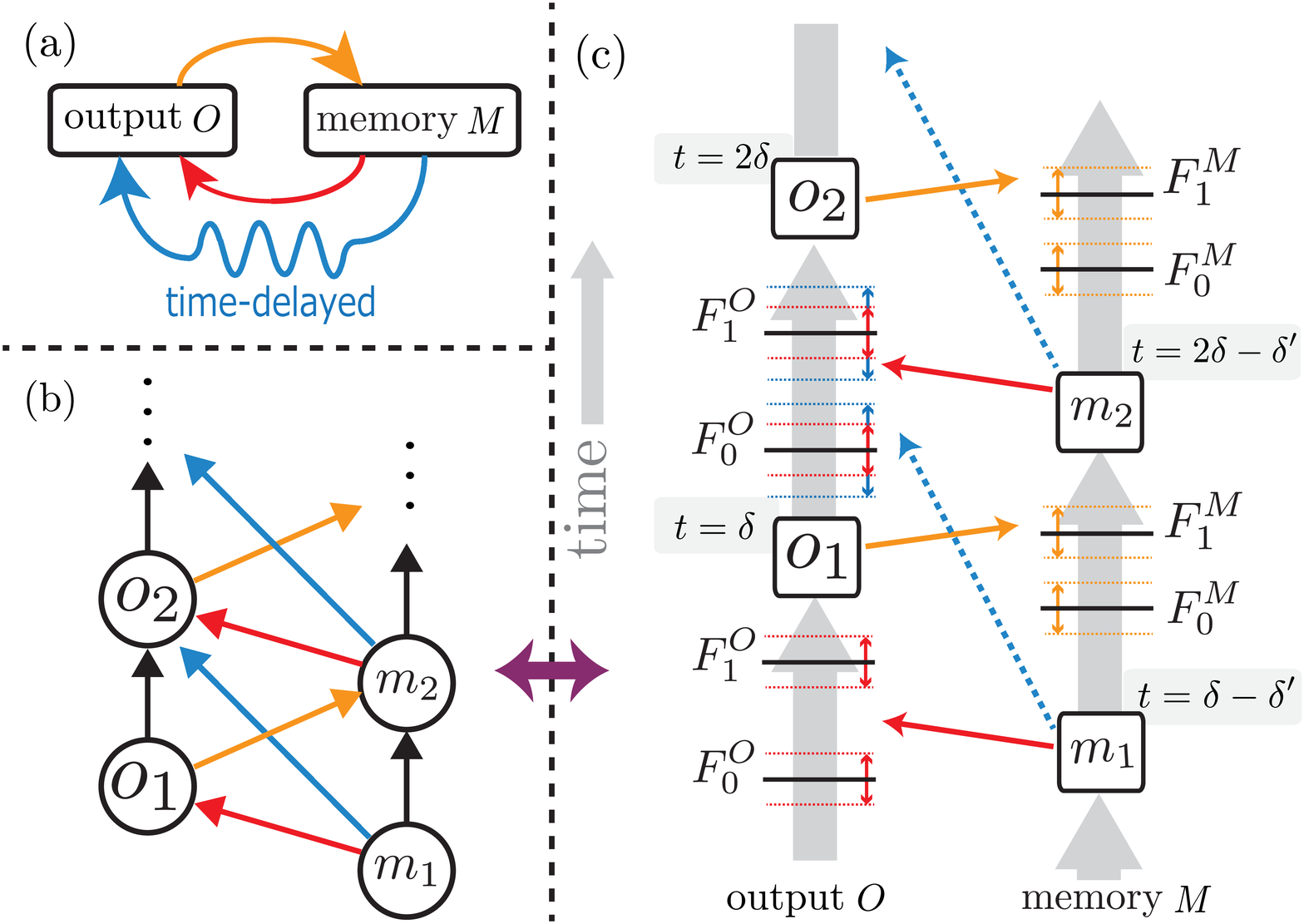}
		\caption{(color online). (a) Feedback loop of a time-delayed chemical reaction model. (b) BN that describes our model. (c) The free-energy levels and the interactions between output $O$ and memory $M$. For instance, $F^{O}_{\mu}(t)$ at time $\delta\leq{t}\leq{2}\delta$ depends on $m_1$ and $m_2$.}
		\label{fig:bayesian3}
	\end{figure}
	The model is characterized by a negative feedback loop between two systems: output system $O$ and memory system $M$ (see Fig.~\ref{fig:bayesian3}~(a)). We assume that each of $O$ and $M$ has a binary state described by $0$ or $1
$. This model is described by the following master equations:
\begin{align}
\frac{dp^{X}_{0}}{dt}(t)\! &=\! - \omega_{0, 1}^{X} (t) p^{X}_{0}(t) + \!\omega_{1, 0}^{X} (t) p^{X}_{1}(t),
\label{master1}
\\
\frac{dp^{X}_{1}}{dt}(t)\! &=\!- \omega_{1, 0}^{X} (t) p^{X}_{1}(t) \!+\! \omega_{0,1}^{X} (t) p^{X}_{0}(t), 
\label{master2}
\end{align}
where $p^{X}_{0}(t)$ and $p^{X}_{1}(t)$ are, respectively, the probabilities of the states $0$ and $1$ with $X=O,M$ at time $t$. 
The transition rate $\omega^{X}_{\mu,\nu}$ ($\mu,\nu=0,1$) is assumed to be
\begin{equation}
\omega^{X}_{\mu,\nu} (t) = \frac{1}{\tau^X} \exp\left\{-\beta^{X}[\Delta^{X}_{\mu \nu} -F^{X}_{\mu}(t)] \right\},
\end{equation}
where $\tau^{X}$ is a time constant, $\beta^{X}$ is the inverse temperature of a heat bath coupled to $X$, $F^{X}_{\mu}(t)$ is the effective free energy of the state $\mu$ at time $t$, $\Delta^{X}_{\mu\nu}$ is a barrier that satisfies $\Delta^{X}_{\mu\nu}=\Delta^{X}_{\nu\mu}$. This transition rate is well established in chemical reaction models~\cite{Sekimoto}. 

Let $o_k$ ($m_{k}$) be the state of $O$ ($M$) at time $t=k\delta$ ($t=k\delta-\delta'$), where $\delta$ is the time interval with $\delta>\delta'$. The feedback loop between $O$ and $M$ is described by $F^{M}_{\mu}(t)$ ($F^{O}_{\mu}(t)$) that depends on $o_k$ ($m_{k}$)  [see also Fig.~\ref{fig:bayesian3}(c)]; we assume that $F^{M}_{\mu}(t)$ depends on $o_k$ at time $k\delta - \delta' \leq t \leq (k+1)\delta -\delta'$, and that $F^{O}_{\mu}(t)$ depends on $m_{k+1}$ and $m_{k}$ at time $k\delta \leq t \leq (k+1)\delta$. The $m_{k}$ dependence of $F^{O}_{\mu}(t)$ describes the effect of time-delayed feedback.

\begin{figure}
		\centering
		\includegraphics[width=90mm]{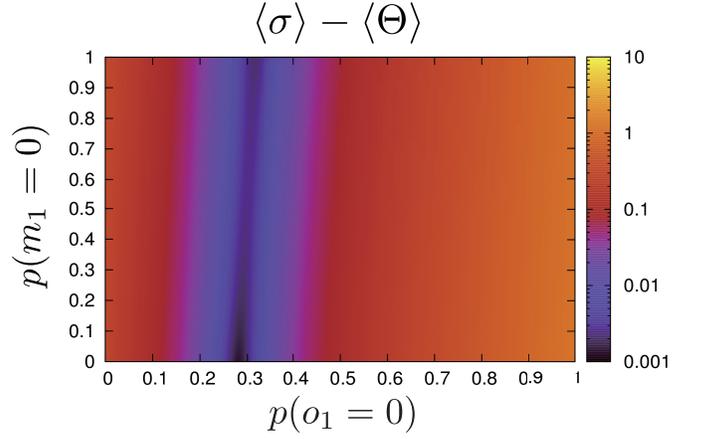}
		\caption{(color online). Numerical illustration of the non-negativity of $\left<\sigma\right>-\left<\Theta\right>=\left<-\beta^{O}Q_{O}\right>+\left<\ln{p}(o_{1},m_{1},m_{2})\right>-\left<\ln{p}(o_{2},m_{1},m_{2})\right>$. We set the initial states to $p(o_1,m_1)=p(o_1)p(m_1)$. The amount of $\left<\sigma\right>-\left<\Theta\right>$ is close to $0$ when the initial states are close to the stationary state of this system. The parameter set is noted in Supplemental Material.}
		\label{fig:bayesian4}
	\end{figure}

By applying Eqs. (\ref{fin mutual information})--(\ref{bayesian information thermodynamics}) to the BN in Fig.~\ref{fig:bayesian3}(b), we obtain two inequalities in the time evolution from $\{o_1,m_1\}$ to $\{o_2,m_2\}$: 
\begin{align}
\left<-\beta^{M} Q_M\right> \geq \left<\ln p(o_{1}, m_{2} )\right> - \left<\ln p(o_{1}, m_{1})  \right>,
\label{2-shanon}
\end{align}
\begin{align}
\left<-\beta^{O} Q_O\right> \geq \left<\ln p(o_{2}, m_{1}, m_{2} )\right> - \left<\ln p(o_{1}, m_{1}, m_{2} )  \right>,
\label{3-shanon}
\end{align}
where $Q_X$ is equal to the effective free-energy difference in this system (see Supplemental Material). The right-hand sides of Eqs. (\ref{2-shanon}) and (\ref{3-shanon}) are the changes in the two-body and three-body Shannon entropies, respectively. This three-body Shannon entropy includes the states of different times $m_1$ and $m_2$. This is a crucial difference between the conventional thermodynamics and our result. We numerically illustrate the validity of Eq. (\ref{3-shanon}) in Fig.~\ref{fig:bayesian4}. We stress that these bounds are calculated from the probability distribution that can be experimentally measured in principle~\cite{Tostevin,Mehta,Cheong}.

\textit{Derivation of the main result.}---From the definition of $\Theta$ in Eqs. (\ref{partition})--(\ref{transfer entropy}), we obtain
\begin{align}
\Theta
\!&=\! \ln \left[ \frac{p(x_N, \mathcal{C})p(x_1)}{p(x_N)p(\mathcal{C})p(x_1|{\rm pa}(x_1))}\prod_{l=1}^{N'} \frac{p(c_l|\mathcal{C}_{l-1})}{p(c_l|{\rm pa}_X(c_l), \mathcal{C}_{l-1})}\right] \nonumber\\
\!&=\! \ln \frac{p(x_1)p(x_N, \mathcal{C})}{p(x_N)p(x_1|{\rm pa}(x_1))\prod_{l=1}^{N'} p(c_l|{\rm pa}(c_l))}\nonumber\\
\!&=\! \ln \frac{p(x_1)p(x_N, \mathcal{C})\prod_{k=2}^{N}p(x_{k}|{\rm pa}(x_{k}))}{p(x_N)p(X, \mathcal{C})}.
\label{calculation}
\end{align}
We then use mathematical properties of BNs~\cite{Bayesian}: $p(c_l|{\rm pa}_{X}(c_l),\mathcal{C}_{l-1})=p(c_l|{\rm pa}(c_l))$ and $p(X,\mathcal{C})\!=\!\prod_{k=1}^{N}\prod_{l=1}^{N'}p(x_k|{\rm pa}(x_k))p(c_l|{\rm pa}(c_l))$ (see the Supplemental Material).
From Eqs. (\ref{definition of joint probability}), (\ref{definition of heat}) and (\ref{calculation}), we arrive at the main result (\ref{bayesian jarzynski equality})
\begin{align}
\left< \exp[-\sigma+ \Theta ] \right>
\!&=\!\sum_{\mathcal{A} }p(\mathcal{D}|\mathcal{C},X)\prod_{k=2}^{N} p_B(x_{k-1}| x_{k}, \mathcal{B}^{k})p(x_N,\mathcal{C})
\nonumber\\
\!&=\!\ 1,
\label{derivation}
\end{align}
where $\mathcal{D}\equiv{\mathcal{A}}\setminus(\mathcal{C}\cup{X})$. Here, we used $\mathcal{B}^{k}\subseteq\mathcal{C}$ ($k=2,\dots,N$) and the normalization of the probability.

\textit{Conclusion.}---In general causal networks, we have derived a novel generalization of the IFT [Eq.~(\ref{bayesian jarzynski equality})].  We have obtained a generalized second law of thermodynamics~(\ref{bayesian information thermodynamics}), which sets a fundamental bound on the entropy production of a single system in the presence of multiple other systems, where the exchanged information between these systems plays a crucial role.

\begin{acknowledgments}
	We are grateful to M. Sano, H. Hayakawa, M. L. Rosinberg, J. M. Horowitz, K. Kanazawa and H. Tasaki	for valuable comments. We acknowledge YITP at Kyoto University. This work was supported by a Grants-in-Aid from JSPS (Grant No. 24$\cdot$8593), by JSPS KAKENHI Grant No. 25800217, and by the Platform for Dynamic Approaches to Living System from MEXT, Japan.
\end{acknowledgments}

\widetext
\appendix
\section{SUPPLEMENTARY INFORMATION}
\section{A: Detailed properties of Bayesian networks}
Let $\mathcal{A}=\{a_j|j=1,2,\dots,N_{\mathcal{A}}\}$ be the set of random variables on a BN, where $a_1,a_2,\dots$ is in the topological ordering. The conditional probability is given by $p(a_j|a_{j-1},\dots, a_1)=p(a_j|{\rm pa}(a_j))$, where ${\rm pa}(a_j) \subseteq \{a_{1}, a_2, \dots , a_{j-1}\}$ is the set of parents of $a_j$.
In this section, we prove two theorems~\cite{Bayesian} that has been used in the derivation of the main result in the main manuscript.

\textbf{Theorem 1 (The chain rule for Bayesian networks).}
\textit{For any index $j$, we have}
\begin{align}
p(a_j, a_{j-1}, \dots, a_1)= \prod_{j'=1}^{j} p(a_{j'}|{\rm pa}(a_{j'})).
\label{The chain rule for Bayesian networks}
\end{align}

\textit{Proof}. 
\begin{align}
p(a_j, a_{j-1}, \dots, a_1) &= p(a_j|a_{j-1}, \dots, a_1)p(a_{j-1}, \dots, a_1) \nonumber\\
&= p(a_j|a_{j-1}, \dots, a_1)p(a_{j-1}| a_{j-2}, \dots, a_1)p(a_{j-2}, \dots, a_1) \nonumber\\
&= \cdots \nonumber\\
&= \prod_{j'=1}^{j} p(a_{j'}|a_{j'-1}, \dots, a_{1} ) \nonumber\\
&= \prod_{j'=1}^{j} p(a_{j'}|{\rm pa}(a_{j'})).\Box
\label{Proof of the chain rule for Bayesian networks}
\end{align}
In the derivation of the main result, we used this theorem as $p(X, \mathcal{C}) = \prod_{k=1}^N \prod_{l=1}^{N'} p(x_k|{\rm pa} (x_k))p(c_l|{\rm pa} (c_l))$, because $\mathcal{C} \cup X = \{c_1, c_2, \dots, c_{N'}, x_1, x_2, \dots, x_N\}=\{a_1, a_2, \dots, a_J \}$, where $a_J$ is chosen to satisfy $a_J = x_N$.

\textbf{Theorem 2 (Consistency of the specification of BN).}
\textit{If $\mathcal{A}'$ is a subset of $\{a_{j-1}, a_{j-2}, \dots, a_1\}$ and ${\rm pa}(a_j)$ is a subset of $\mathcal{A}'$ (${\rm pa}(a_j) \subseteq \mathcal{A}' \subseteq \{a_{j-1},a_{j-2},  \dots, a_1\}$), we have}
\begin{align}
p(a_j|\mathcal{A}')= p(a_j|{\rm pa}(a_j)).
\label{Consistency of the specification of BN}
\end{align}

\textit{Proof}. 
\begin{align}
p(a_j|\mathcal{A}') &= \frac{p(a_j, \mathcal{A}')}{p(\mathcal{A}')} \nonumber\\
&= \frac{\sum_{\{a_{j},a_{j-1},  \dots, a_1\} \setminus \{a_j, \mathcal{A}' \}}p(a_j, a_{j-1}, \dots, a_1 )}{\sum_{\{a_{j},a_{j-1},  \dots, a_1\} \setminus \{\mathcal{A}' \}}p(a_j, a_{j-1}, \dots, a_1 )}\nonumber\\
&= \frac{\sum_{\{a_{j},a_{j-1},  \dots, a_1\} \setminus \{a_j,\mathcal{A}' \}}\prod_{j'=1}^{j} p(a_{j'}|{\rm pa}(a_{j'}))}{\sum_{\{a_{j-1},  \dots, a_1\} \setminus \{\mathcal{A}' \}}p(a_{j-1},a_{j-2}, \dots, a_1 )}\nonumber \nonumber\\
&= p(a_{j}|{\rm pa}(a_{j}))\frac{\sum_{\{a_{j-1},  \dots, a_1\} \setminus \{\mathcal{A}' \}}\prod_{j'=1}^{j-1} p(a_{j'}|{\rm pa}(a_{j'}))}{\sum_{\{a_{j-1},  \dots, a_1\} \setminus \{\mathcal{A}' \}}p(a_{j-1},a_{j-2}, \dots, a_1 )}\nonumber\\
&= p(a_{j}|{\rm pa}(a_{j})).\Box
\label{Proof of consistency of the specification of BN}
\end{align}
In the derivation of the main result, we used this theorem as $p(c_l|{\rm pa}_X(c_l), \mathcal{C}_{l-1}) = p(c_l|{\rm pa}(c_l)) $, because ${\rm pa}(a_{J'}) \subseteq \{{\rm pa}_X(c_l), \mathcal{C}_{l-1}\} \subseteq \{a_{J'-1},a_{J'-2},  \dots, a_1\}$, where $a_{J'}$ is chosen to satisfy $a_{J'}=c_l$.

\section{B: Physical meaning of the transfer entropy}
In Ref.~\cite{Schreiber}, Schreiber introduced the transfer entropy for stochastic dynamics with two variables $I=\{ i_1, i_2, \dots,i_n, \dots\}$ and $J=\{ j_1, j_2, \dots,j_n,\dots \}$, where $i_n$ ($j_n$) denotes the state of the system $I$ ($J$) at time $n$. The transfer entropy from $J$ to $I$ is defined by
\begin{align}
T_{J\to I}\equiv \sum p(I,J) \ln \frac{p(i_{n+1}, j_{n}, j_{n-1}, \dots, j_{n-l} |i_{n}, i_{n-1} \dots, i_{n-k})}{p(i_{n+1}|i_{n}, i_{n-1} \dots, i_{n-k}) p(j_{n}, j_{n-1}, \dots, j_{n-l} |i_{n}, i_{n-1} \dots, i_{n-k})},
\end{align}
Here, $T_{I\to J}$ characterizes the information flow from $I$ to $J$; in fact, $T_{I\to J}$ is given by the difference between the entropy rate in $I$ and that under the condition of $J$:
\begin{align}
T_{J\to I}\equiv \Delta s_{I|J} - \Delta s_{I},
\end{align}
where the entropy rate in $I$ and that under the condition of $J$, $\Delta s_{I}$ and $\Delta s_{I|J}$, are defined as $\Delta s_{I} \equiv  \sum p(I)[\ln p(i_{n+1}, i_{n}, i_{n-1} \dots, i_{n-k})- \ln p(i_{n}, i_{n-1} \dots, i_{n-k})]$ and $\Delta s_{I|J} \equiv \sum p(I,J)[ \ln p(i_{n+1}, i_{n}, i_{n-1} \dots, i_{n-k}| j_{n}, j_{n-1}, \dots, j_{n-l})- \ln p(i_{n}, i_{n-1} \dots, i_{n-k}| j_{n}, j_{n-1}, \dots, j_{n-l})]$, respectively.

The quantity $\left< I_{\rm tr}^l \right>$ in our main result is defined as
\begin{align}
\left< I_{\rm tr}^l\right> \equiv  \sum_{\mathcal{C}, X} p(\mathcal{C}, X) \ln \frac{p(c_l, {\rm pa}_X(c_l)| c_{l-1}, \dots, c_1 )}{p(c_l| c_{l-1}, \dots, c_1 )p({\rm pa}_X(c_l)| c_{l-1}, \dots, c_1 )},
\end{align}
which equals the transfer entropy $T_{X\to \mathcal{C}}$.
\section{C: Multidimensional Langevin systems}
We consider the following multidimensional over-damped Langevin equation:
\begin{align}
\gamma^{(\mu')} \dot{x}^{(\mu')}(t) \!=\! f^{(\mu')} (x^{(1)}(t), \dots , x^{(n')}(t)) +\xi^{(\mu')} (t), 
\label{Langevin1}
\end{align}
\begin{align}
\left< \xi^{(\mu')}(t) \xi^{(\nu')}(t') \right> &= 2 \gamma^{(\mu')} k_B T^{(\mu')} \delta_{\mu' \nu'} \delta(t-t')
\label{Langevin2}\\
\left< \xi^{(\mu')}(t)\right> &= 0,
\label{Langevin3}
\end{align}
where $x^{(\mu')}$ ($\mu' =1, \dots, n'$) denotes a dynamical variable. With small time interval $\Delta t$, we discretize the dynamical variables as $x^{(\mu')}_k\equiv x^{(\mu')}(k\Delta t)$. We write $x^{(1)}_k\equiv x_k$. When ${\bm x}_{k}\equiv \{ x^{(2)}_k, \dots, x^{(n')}_k \}$ is fixed, we obtain the conditional probability $p(x_{k+1}|x_k, {\bm x}_k)$ in terms of the Stratonovich product:
\begin{align}
p(x_{k+1}|x_k, {\bm x}_k)=\mathcal{N} \exp \left[-\frac{\Delta t}{4\gamma^{(1)} k_{B} T^{(1)}} \left( \gamma^{(1)} \frac{\epsilon^{(1)}_k}{\Delta t}- f^{(1)} (\bar{x}^{(1)}_k , {\bm x}_k) \right)^2 - \frac{\Delta t}{2}\frac{\partial}{\partial x^{(1)}} f^{(1)} (\bar{x}^{(1)}_k, {\bm x}_k) \right],
\label{forwardpath}
\end{align}
where $\bar{x}^{(\mu')}_k \equiv (x^{(\mu')}_k + x^{(\mu')}_{k+1})/2$, $\epsilon^{(\mu')}_k \equiv x^{(\mu')}_{k+1} -x^{(\mu')}_{k}$, $f^{(1)} (\bar{x}_k, {\bm x}_k) \equiv f^{(1)} (\bar{x}_k, x^{(2)}_k, \dots, x^{(n')}_k)$, and $\mathcal{N}$ is the prefactor which does not depend on $f^{(\mu')}$~\cite{Chernyak}. We stress that we use the mid-point rule only for $x^{(1)}$. We define the conditional probability of the backward process $p_B(x_{k}|x_{k+1}, {\bm x}_{k})$ as
\begin{align}
p_B(x_{k}|x_{k+1}, {\bm x}_{k})=\mathcal{N} \exp \left[-\frac{\Delta t}{4\gamma^{(1)} k_{B} T^{(1)}} \left( -\gamma^{(1)} \frac{\epsilon^{(1)}_k}{\Delta t}- f^{(1)} (\bar{x}^{(1)}_k , {\bm x}_k) \right)^2 - \frac{\Delta t}{2}\frac{\partial}{\partial x^{(1)}} f^{(1)} (\bar{x}^{(1)}_k, {\bm x}_k) \right].
\label{backwardpath}
\end{align}
			\begin{figure}
			\centering
			\includegraphics[width=80mm,clip]{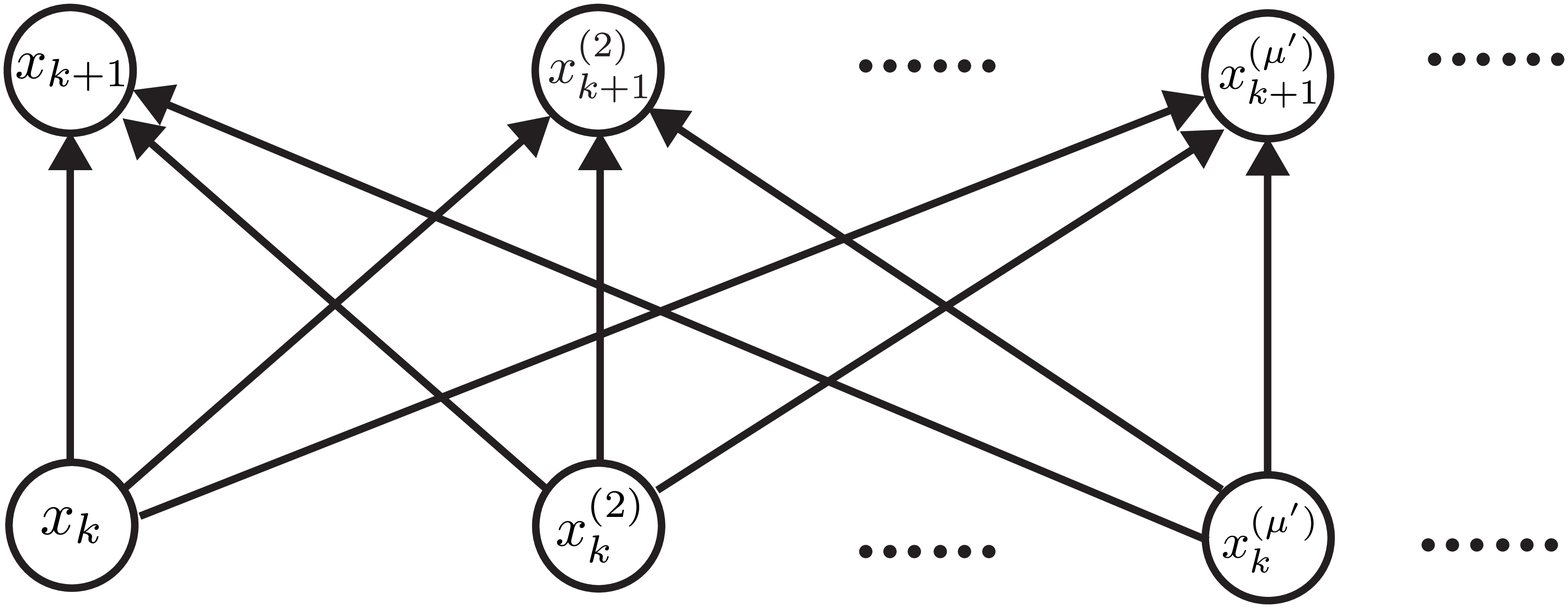}
			\caption{BN corresponding to the multidimensional Langevin equation.
					}
			\label{figsupple:langevin}
		\end{figure}
Figure~\ref{figsupple:langevin} shows the Bayesian network (BN) corresponding to the multidimensional Langevin equation [Eqs. (\ref{Langevin1}), (\ref{Langevin2}) and (\ref{Langevin3})] for the time interval between $k\Delta t$ and $(k+1) \Delta t$. Thus we have $\mathcal{B}^{k+1} = {\bm x}_k$. From Eqs. (\ref{forwardpath}) and (\ref{backwardpath}) , we obtain $\Delta s_{\rm bath}^{k+1}$:
\begin{align}
\Delta s_{\rm bath}^{k+1} &= \ln \frac{p(x_{k+1}|x_k, {\bm x}_k)}{p_B(x_{k}|x_{k+1}, {\bm x}_{k})}\\
&= -\frac{1}{k_B T^{(1)}} f^{(1)} (\bar{x}^{(1)}_k , {\bm x}_k) \epsilon^{(1)}_k.
\end{align}
The definition of the heat flux in system $x^{(1)}$ by Sekimoto~\cite{Sekimoto2, Sekimoto3} is given by $J^{(1)} \equiv f^{(1)} (\bar{x}^{(1)}_k, \bar{x}^{(2)}_k, \dots, \bar{x}^{(n')}_k) \epsilon^{(1)}_k$. We then compare $\Delta s'_{\rm bath} \equiv - J^{(1)}/ (k_{B} T^{(1)})$ with $\Delta s_{\rm bath}^{k+1}$ as
\begin{align}
\Delta s'_{\rm bath} -\Delta s_{\rm bath}^{k+1}  &= - \frac{1 }{k_{B} T^{(1)}} \left[\sum_{\mu' =2}^{n'} \frac{\partial f^{(1)}} {\partial x^{(\mu')}} (\bar{x}^{(1)}_k, x^{(2)}_k, \dots , x^{(\mu')}_k,  \bar{x}^{(\mu'+1)}_k, \dots, \bar{x}^{(n')}_k) \epsilon^{(1)}_k \epsilon^{(\mu')}_k \right] \\
&= o(\Delta t ),
\end{align}
where we used $\epsilon^{(1)}_k \epsilon^{(\mu')}_k = o(\Delta t)$ with $\mu' \neq 1$ because of the independence of the noises [Eq. (\ref{Langevin2})].
Therefore, our definition of the entropy change in the heat baths on the BN (\textit{i.e.}, $\Delta s_{\rm bath}^{k+1}$) is equivalent to the Sekimoto's definition (\textit{i.e.}, $\Delta s'_{\rm bath}$) up to $o(\Delta t)$.
\section{D: Repeated feedback control}
\begin{figure}
			\centering
			\includegraphics[width=50mm,clip]{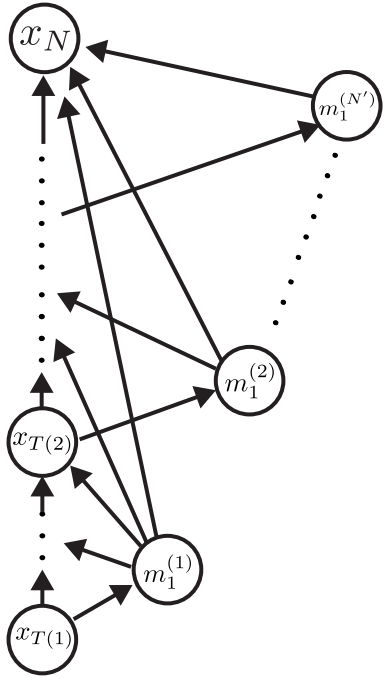}
			\caption{BN corresponding to the repeated feedback control.
					}
			\label{figsupple:repeated}
		\end{figure}

	We consider systems under repeated feedback control. Figure~\ref{figsupple:repeated} shows the BN corresponding to the repeated feedback control discussed by Horowitz and Vaikuntanathan~\cite{Horowitz}.
	There are system $X$ and memories $M^{(\mu')}$ with $\mu'=1,\dots ,N'$ ($N' \leq N$). Measurements are performed on system $X$ at time $T(\mu')$, where $T(\mu')$ is the natural number such as $1=T(1)<T(2)<\cdots<T(N')<N$. The state of $X$ at time $T(\mu')$ is given by $x_{T(\mu')}$, where the measurement outcome is stored in $m^{(\mu')}_1$. The states of $X$ under feedback control can then depend on $m^{(\mu')}_1$ after time $T(\mu')$.

	We have $\mathcal{C} = \{m^{(1)}_1, m^{(2)}_1 ,\dots ,m^{(N')}_1 \}$ and ${\rm pa}(x_1) =\emptyset$, and therefore $I_{\rm fin}= I(x_{N}: \{m^{(1)}_1, \dots , m^{(N')}_1\})$, $I_{\rm ini} = 0$, $I^{l}_{\rm tr}= I(x_{T(l)}: m^{(l)}_1|m^{(l-1)}_1 , \dots , m^{(1)}_{1})$, and
\begin{equation}	
\Theta= I(x_{N}: \{m^{(1)}_1, \dots , m^{(N')}_1\})-\sum_{l=1}^{N'} I(x_{T(l)}: m^{(l)}_1|m^{(l-1)}_1 , \dots , m^{(1)}_{1}).
\end{equation}
	According to the main result, we obtain the following generalized Jarzynski equality:
\begin{equation}	
\left<\exp[-\sigma+ \Theta ]\right>= 1.
\label{bayesian jarzynski}
\end{equation}

	On the other hand, the equality derived by Horowitz and Vaikuntanathan~\cite{Horowitz} is given by
\begin{equation}	
\left<\exp \left[-\beta W_d-\sum_{l=1}^{N'} I^{l}_{\rm tr} \right] \right> =1,
\label{Horowitz jarzynski}
\end{equation}
where $I_{\rm tr}^l$ is our definition of the transfer entropy that is given by $I^{l}_{\rm tr}= I(x_{T(l)}: m^{(l)}_1|m^{(l-1)}_1 , \dots , m^{(1)}_{1})$ and $\beta$ is the inverse temperature of the heat bath. $W_d$ is the dissipated work that is given by $\beta W_d \equiv \Delta s_{\rm bath} + \ln p_{\rm eq}(x_1)-\ln p_{\rm eq}(x_N|m_1^{(1)} , \dots, m_{1}^{(N')})$, where $p_{\rm eq}$ is the canonical equilibrium distribution for fixed control parameter. $\beta W_d$ is equivalent to $\sigma-I_{\rm fin}$ such that $\sigma - I_{\rm fin}= \Delta s_{\rm bath} + \ln p(x_1)- \ln p(x_N|m_1^{(1)} , \dots, m_{1}^{(N')})$.

Therefore, our result can reproduce the result obtained by Horowitz and Vaikuntanathan in Ref.~\cite{Horowitz}, when the initial and final states of the system are in thermal equilibrium.

\section{E: Detailed calculations in the adaptation model}
	In the adaptation model in the main manuscript, we consider the following master equations:
\begin{align}
\frac{dp^{X}_{0}}{dt} (t)\! &=\! - \omega_{0, 1}^{X}(F^X_{0}(t) ) p^{X}_{0} (t)+ \!\omega_{1, 0}^{X} (F^X_{1} (t)) p^{X}_{1}(t), 
\label{mastersup1}
\\
\frac{dp^{X}_{1}}{dt} (t)\! &=\!- \omega_{1, 0}^{X}(F^X_{1}(t)) p^{X}_{1} (t)\!+\! \omega_{0,1}^{X}(F^X_{0}(t)) p^{X}_{0}(t).
\label{mastersup2}
\end{align}
where the transition rate is given by
\begin{equation}
\omega^{X}_{\mu,\nu} (F^X_{\mu}(t))= \frac{1}{\tau^X} \exp\left[-\beta^{X}(\Delta^{X}_{\mu \nu} -F^{X}_{\mu}(t)) \right].
\label{transition rate}
\end{equation}
In the following, we show that $\Delta s_{\rm bath}^{k+1}$ is equal to $-\beta^X \Delta F^{X}$.

We note that $p^{X}_{0}(t)+p^{X}_{1}(t)=1$ holds because of the normalization of the probability distribution.
We rewrite Eq. (\ref{mastersup1}) as
\begin{align}
\frac{dp^{X}_{0}}{dt} (t)\! &=\! - [ \omega_{0, 1}^{X}(F^X_{0}(t) )  + \omega_{1, 0}^{X}(F^X_{1} (t)  )] p^{X}_{0}(t) + \omega_{1, 0}^{X} (F^X_{1}(t)).
\label{Master}
\end{align}
When $F^X_{0}$ and $F^X_{1}$ are constants, we get the solution of Eq. (\ref{Master}) as
\begin{align}
p^{X}_{0}(t) =p^{X}_{0,{\rm eq}}+  (p^{X}_{0} (0) - p^{X}_{0,{\rm eq}} ) \exp \left[ - (\omega_{0, 1}^{X}(F^X_{0}) + \omega_{1, 0}^{X} (F^X_{1}))t \right],
\label{solution}
\end{align}
where $p^{X}_{0,{\rm eq}}$ is defined as
\begin{align}
p^{X}_{0,{\rm eq}}(F^{X}_0, F^{X}_1) \equiv \frac{\omega_{1, 0}^{X}(F^X_{1})}{\omega_{0, 1}^{X}(F^X_{0}) +  \omega_{1, 0}^{X}(F^X_{1})} =\frac{\exp(-\beta^X F^X_{0} )}{\exp(-\beta^X F^X_{0})+\exp(-\beta^X F^X_{1})}.
\end{align}

The state of $O$ ($M$) at time $t=k\delta$ ($t=k\delta - \delta'$) describes $o_k$ ($m_{k}$) with $\delta > \delta'$. We set the interaction between the memory $X=M$ and the output system $X=O$ as follow. Let $F^{M}_{\mu}(t)$ at time $k\delta - \delta' \leq t \leq (k+1)\delta - \delta'$ be
\begin{eqnarray}
F^{M}_{\mu}(t) =  \begin{array}{ll} F_{\mu,j'} & (o_{k}=j'), \end{array}
\end{eqnarray}
and let $F^{O}_{\mu}(t)$ at time $k\delta \leq t \leq (k+1)\delta$ be
\begin{eqnarray}
F^{O}_{\mu}(t) = \begin{array}{ll}
    F'_{\mu,j'k'} & (m_{k}=j',m_{k+1}=k'),
  \end{array} 
  \end{eqnarray}
where $j', k' =0 ,1$.

Substituting $p_0^{M}(0)=0,1$ into the solution of Eq. (\ref{solution}), we have the conditional probabilities $p(m_{k+1}|m_{k},o_{k})$:
\begin{eqnarray}
p(m_{k+1}=0|m_{k}=0,o_k=j') &=& q_{j'}+ \left(1 - q_{j'}\right)\exp \left[ -\omega_{j'} \delta \right], 
\label{so1}\\
p(m_{k+1}=0|m_{k}=1,o_k=j') &=& q_{j'} - q_{j'}\exp \left[ -\omega_{j'}  \delta \right], 
\label{so2}\\
p(m_{k+1}=1|m_{k}=i',o_k=j') &=& 1- p(m_{k+1}=0|m_{k}=i',o_k=j') ,
\label{so3}
\end{eqnarray}
where $i'=0,1$, $q_{j'} \equiv p_{0,\rm eq}^{M} (F_{0,j'}, F_{1,j'})$ and $\omega_{j'} \equiv \omega_{0, 1}^{M}(F_{0,j'}) + \omega_{1, 0}^{M} (F_{1,j'})$. 
Substituting $p_0^{O}(0)=0,1$ into Eq. (\ref{solution}), we also have the conditional probabilities $p(o_{k+1}|o_{k},m_{k},m_{k+1})$:
\begin{eqnarray}
p(o_{k+1}=0|o_{k}=0,m_{k}=j' ,m_{k+1}=k') &=& q'_{j'k'}+ \left(1 - q'_{j'k'} \right)\exp \left[ -\omega'_{j'k'} \delta \right] 
\label{so4}\\
p(o_{k+1}=0|o_{k}=1,m_{k}=j' ,m_{k+1}=k') &=& q'_{j'k'} - q'_{j'k'} \exp \left[ -\omega'_{j'k'} \delta \right] 
\label{so5}\\
p(o_{k+1}=1|o_{k}=i',m_{k}=j' ,m_{k+1}=k')&=&1-  p(o_{k+1}=0|o_{k}=i' ,m_{k}=j' ,m_{k+1}=k'),
\label{so6}
\end{eqnarray}
where $q'_{j'k'} \equiv p_{0,\rm eq}^{O} (F'_{0,j'k'}, F'_{1,j'k'})$ and $\omega'_{j'k'} \equiv \omega_{0, 1}^{O}(F'_{0,j'k'}) + \omega_{1, 0}^{O} (F'_{1,j'k'})$.

We assume that the conditional probabilities of backward process $p_B(m_{k}|m_{k+1},o_{k})$ and $p_B(o_{k}|o_{k+1},m_{k},m_{k+1})$ are defined as $p_B(m_{k}=l'|m_{k+1}=i',o_{k}=j') \equiv p(m_{k+1}=l'|m_{k}=i',o_{k}=j')$ and $p_B(o_{k}=l'|o_{k+1}=i',m_{k}=j',m_{k+1}=k') \equiv p(o_{k+1}=l'|o_{k}=i',m_{k}=j',m_{k+1}=k')$ with $l'=0,1$, respectively.

From Eqs. (\ref{transition rate}), (\ref{so1}), (\ref{so2}) and (\ref{so3}), we have $\Delta s_{\rm bath}^{k+1}$ with $X=M$:
\begin{align}
\Delta s_{\rm bath}^{k+1} &=\ln \left[ \frac{p(m_{k+1}|m_{k},o_{k}) }{p_B(m_{k}|m_{k+1},o_{k})  } \right] \\
&=\left\{ \begin{array}{ll}
    0  & (m_{k+1}=0, m_{k}=0, o_{k}=j'))\\
    \ln q_{j'} -\ln(1-q_{j'}) & (m_{k+1}=0, m_{k}=1, o_{k}=j') \\
     \ln (1- q_{j'}) -\ln q_{j'} & (m_{k+1}=1, m_{k}=0, o_{k}=j')\\
    0  & (m_{k+1}=1, m_{k}=1, o_{k}=j')\\
  \end{array} \right. \\
&=\begin{array}{ll} -\beta^M( F_{l',j'} -F_{i',j'}) & (m_{k+1}=l', m_{k}=i', o_{k}=j'),  \end{array}
\end{align}
where $\mathcal{B}^{k+1}= \{o_{k} \}$. From Eqs. (\ref{transition rate}), (\ref{so4}), (\ref{so5}) and (\ref{so6}), we have $\Delta s_{\rm bath}^{k+1}$ with $X=O$:
\begin{align}
\Delta s_{\rm bath}^{k+1} &=\ln \left[ \frac{p(o_{k+1}|o_{k},m_{k}, m_{k+1}) }{p_B(o_{k}|o_{k+1},m_{k}, m_{k+1})  } \right] \\
&=\left\{ \begin{array}{ll}
    0  & (o_{k+1}=0, o_{k}=0, m_{k}=j', m_{k+1}=k' )\\
    \ln q'_{j'k'} -\ln(1-q'_{j'k'}) & (o_{k+1}=0, o_{k}=1, m_{k}=j', m_{k+1}=k' )\\
     \ln (1- q'_{j'k'}) -\ln q'_{j'k'} & (o_{k+1}=1, o_{k}=0, m_{k}=j', m_{k+1}=k' )\\
    0  & (o_{k+1}=1, o_{k}=1, m_{k}=j', m_{k+1}=k' )\\
  \end{array} \right. \\
&=\begin{array}{ll} -\beta^O( F'_{l',j'k'} -F'_{i',j'k'}) & (o_{k+1}=l' ,o_{k}=i', m_{k}=j', m_{k+1}=k'),  \end{array}
\end{align}
where $\mathcal{B}^{k+1}= \{m_{k}, m_{k+1} \}$, we reach the conclusion that $\Delta s_{\rm bath}^{k+1}$ is given by the effective free-energy difference.

\section{F: The parameter set of the numerical illustration in the adaptation model}
We set the parameters of the numerical illustration in Fig.~5 of the main manuscript as follows: $\delta=0.5$, $\beta^M = \beta^O =0.01$, $\tau^O =\tau^M =0.001$, $\Delta^{M}_{01}= \Delta^O_{01}=100$, $F_{0,0}=F_{0,1} = 100$, $F_{1,0} =10$, $F_{1,1}=30$, $F'_{0,00}=F'_{0,01}=F'_{0,10}=F'_{0,11}=100$, $F'_{1,00}=30$, $F'_{1,01}=20$, $F'_{1,10}=10$ and $F'_{1,11}=5$. In this case, we have $q'_{00}=0.332$, $q'_{01}=0.310$, $q'_{10}=0.289$ and $q'_{11}=0.278$. We note that the value of $\left<  \sigma \right>- \left< \Theta \right>$ in Fig.~5 of the main manuscript is close to $0$ when the initial states are close to the stationary distribution of the output system, which is similar to the probabilities $q'_{00}$, $q'_{01}$, $q'_{10}$ and $q'_{11}$.

\end{document}